\documentclass{aastex631}

\usepackage{xspace}
\newcommand{\galfit}{{\scshape Galfit}\xspace}
\newcommand{\galfitm}{{\scshape Galfitm}\xspace}

\newcommand{\daophot}{{\scshape DAOphot}\xspace}
\newcommand{\ESOeclipse}{{\scshape ESO-eclipse}\xspace}
\newcommand{\jitter}{{\scshape jitter}\xspace}
\newcommand{\clic}{{\scshape CLIC}\xspace}
\newcommand{\mapping}{{\scshape MAPPING}\xspace}
\newcommand{\gildas}{{\scshape GILDAS}\xspace}

\newcommand{\sersic}{S\'ersic\xspace}
\newcommand{\lmass}{\hbox{$\log(\mathrm{M}_{\ast} / \mathrm{M}_{\odot})$}}
\newcommand{\re}{\hbox{$R_{\rm{e}}$}}

\newcommand{\hst}{{\em HST}}
\newcommand{\jwst}{{\em JWST}}

\begin{document}

\title{The first {\em JWST} spectrum of a GRB afterglow: No bright supernova in observations of the brightest GRB of all time, GRB 221009A}

\author[0000-0001-7821-9369]{A. J. Levan}
\affiliation{Department of Astrophysics/IMAPP, Radboud University, 6525 AJ Nijmegen, The Netherlands.}
\affiliation{Department of Physics, University of Warwick, Coventry, CV4 7AL, UK.}

\author[0000-0001-5169-4143]{G. P. Lamb}
\affiliation{Astrophysics Research Institute, Liverpool John Moores University, 146 Brownlow Hill, Liverpool L3 5RF, UK.}

\author[0000-0003-4876-7756]{B. Schneider}
\affiliation{Kavli Institute for Astrophysics and Space Research, Massachusetts Institute of Technology, 77 Massachusetts Ave, Cambridge, MA 02139, USA.}

\author[0000-0002-4571-2306]{J. Hjorth}
\affiliation{DARK, Niels Bohr Institute, University of Copenhagen, Jagtvej 128, 2200 Copenhagen, Denmark.}

\author[0000-0002-0786-7307]{T. Zafar}
\affiliation{Australian Astronomical Optics, Macquarie University, 105 Delhi Road, North Ryde, NSW 2113, Australia.}

\author[0000-0002-0786-7307]{A. de Ugarte Postigo}
\affiliation{Artemis, Universit\'e C\^ote d'Azur, Observatoire de la C\^ote d'Azur, CNRS, F-06304 Nice, France.}

\author[0000-0001-9855-8261] {B. Sargent}
\affiliation{Space Telescope Science Institute, 3700 San Martin Drive, Baltimore, MD 21218, USA.}
\affiliation{Center for Astrophysical Sciences, The William H. Miller III Department of Physics 
and Astronomy, Johns Hopkins University, Baltimore, Maryland 21218, USA.}

\author[0000-0001-7106-4683]{S. E. Mullally}
\affiliation{Space Telescope Science Institute, 3700 San Martin Drive, Baltimore, MD 21218, USA.}

\author[0000-0001-9695-8472]{L. Izzo}
\affiliation{DARK, Niels Bohr Institute, University of Copenhagen, Jagtvej 128, 2200 Copenhagen, Denmark.}

\author[0000-0001-7164-1508]{P. D'Avanzo}
\affiliation{INAF, Osservatorio Astronomico di Brera, Via E. Bianchi 46, I-23807, Merate (LC), Italy.}

\author[0000-0002-2942-3379]{E. Burns}
\affiliation{Department of Physics and Astronomy, Louisiana State University, Baton Rouge, LA 70803 USA.}

\author[0000-0001-6991-7616]{J. F. Ag\"u\'i Fern\'andez}
\affiliation{Instituto de Astrof\'isica de Andaluc\'ia - CSIC, Glorieta de la Astronom\'ia s/n, 18008 Granada, Spain.}

\author[0000-0001-7139-2724]{T. Barclay}
\affiliation{NASA Goddard Space Flight Center, Greenbelt, MD 20771, USA.}
\affiliation{University of Maryland, Baltimore County, Baltimore, MD 21250, USA.}

\author[0000-0001-6106-3046]{M. G. Bernardini}
\affiliation{INAF, Osservatorio Astronomico di Brera, Via E. Bianchi 46, I-23807, Merate (LC), Italy.}

\author[0000-0003-0136-1281]{K. Bhirombhakdi}
\affiliation{Space Telescope Science Institute, 3700 San Martin Drive, Baltimore, MD 21218, USA.}

\author[0000-0001-7511-3745]{M. Bremer} 
\affiliation{Institut de Radioastronomie Millim\'etrique (IRAM), 300 Rue de la Piscine, F-38406 Saint Martin d'H\`eres, France.}

\author{R. Brivio}
\affiliation{Universit\'a degli Studi dell’Insubria, Dipartimento di Scienza e Alta Tecnologia, via Valleggio 11, 22100 Como, Italy.}
\affiliation{INAF, Osservatorio Astronomico di Brera, Via E. Bianchi 46, I-23807, Merate (LC), Italy.}

\author[0000-0001-6278-1576]{S. Campana}
\affiliation{INAF, Osservatorio Astronomico di Brera, Via E. Bianchi 46, I-23807, Merate (LC), Italy.}

\author[0000-0001-9842-6808]{A. A. Chrimes}
\affiliation{Department of Astrophysics/IMAPP, Radboud University, 6525 AJ Nijmegen, The Netherlands.}

\author[0000-0002-7320-5862]{V. D'Elia}
\affiliation{ASI, Italian Space Agency, Space Science Data Centre, Via del Politecnico snc, 00133 Rome, Italy.}
\affiliation{INAF, Osservatorio Astronomico di Roma, Via Frascati 33, I-00040 Monte Porzio Catone (RM), Italy.}

\author[0000-0003-3142-5020]{M. Della Valle}
\affiliation{Capodimonte Astronomical Observatory, INAF-Napoli, Salita Moiariello 16, I-80131 Napoli, Italy}

\author[0000-0002-4036-7419]{M. De Pasquale}
\affiliation{University of Messina, Mathematics, Informatics, Physics and Earth Science Department, Via F.D. D'Alcontres 31, Polo Papardo, 98166, Messina, Italy.}

\author{M. Ferro}
\affiliation{Universit\'a degli Studi dell’Insubria, Dipartimento di Scienza e Alta Tecnologia, via Valleggio 11, 22100 Como, Italy.}
\affiliation{INAF, Osservatorio Astronomico di Brera, Via E. Bianchi 46, I-23807, Merate (LC), Italy.}

\author[0000-0002-7374-935X]{W. Fong}
\affiliation{Center for Interdisciplinary Exploration and Research in Astrophysics (CIERA) and Department of Physics and Astronomy, Northwestern University, Evanston, IL 60208, USA.}

\author[0000-0002-6652-9279] {A. S. Fruchter}
\affiliation{Space Telescope Science Institute, 3700 San Martin Drive, Baltimore, MD 21218, USA.}

\author[0000-0003-3457-9375]{J. P. U. Fynbo}
\affiliation{The Cosmic Dawn Center.}
\affiliation{Niels Bohr Institute, Copenhagen University, Jagtvej 155, DK-2200, Copenhagen N, Denmark.}

\author[0000-0002-3855-707X]{N. Gaspari}
\affiliation{Department of Astrophysics/IMAPP, Radboud University, 6525 AJ Nijmegen, The Netherlands.}

\author[0000-0002-5826-0548]{B. P. Gompertz}
\affiliation{School of Physics and Astronomy \& Institute for Gravitational Wave Astronomy, University of Birmingham, Birmingham B15 2TT, UK.}

\author[0000-0002-8028-0991]{D. H. Hartmann}
\affiliation{Department of Physics and Astronomy, Clemson University, Clemson, SC 29634, USA.}

\author{C. L. Hedges}
\affiliation{NASA Goddard Space Flight Center, Greenbelt, MD 20771, USA.}
\affiliation{University of Maryland, Baltimore County, Baltimore, MD 21250, USA.}

\author[0000-0002-9389-7413]{K. E. Heintz}
\affiliation{The Cosmic DAWN Center.}
\affiliation{Niels Bohr Institute, Copenhagen University, Jagtvej 155, DK-2200, Copenhagen N, Denmark.}

\author[0000-0002-2502-3730]{K. Hotokezaka}
\affiliation{Research Center for the Early Universe, Graduate School of Science, University of Tokyo, Bunkyo, Tokyo 113-0033, Japan.}

\author[0000-0002-9404-5650]{P. Jakobsson}
\affiliation{Centre for Astrophysics and Cosmology, Science Institute, University of Iceland, Dunhagi 5, 107 Reykjav\'ik, Iceland.}

\author[0000-0003-2902-3583]{D. A. Kann}
\affiliation{Hessian Research Cluster ELEMENTS, Giersch Science Center, Max-von-Laue-Stra{\ss}e 12, Goethe University Frankfurt, Campus Riedberg, 60438 Frankfurt am Main, Germany.}

\author[0000-0002-6745-4790]{J. A. Kennea}
\affiliation{Department of Astronomy and Astrophysics, Pennsylvania State University, 525 Davey Lab, University Park, PA 16802, USA.}

\author[0000-0003-1792-2338]{T. Laskar}
\affiliation{Department of Physics \& Astronomy, University of Utah, Salt Lake City, UT 84112, USA.}
\affiliation{Department of Astrophysics/IMAPP, Radboud University, 6525 AJ Nijmegen, The Netherlands.}

\author[0000-0001-7421-4413]{E. Le Floc'h}
\affiliation{Universit\'e Paris-Saclay, Universit\'e Paris Cité, CEA, CNRS, AIM, 91191, Gif-sur-Yvette, France.}

\author[0000-0002-7517-326X]{D. B. Malesani}
\affiliation{Department of Astrophysics/IMAPP, Radboud University, 6525 AJ Nijmegen, The Netherlands.}
\affiliation{The Cosmic DAWN Center.}
\affiliation{Niels Bohr Institute, Copenhagen University, Jagtvej 155, DK-2200, Copenhagen N, Denmark.}

\author[0000-0002-2810-2143]{A. Melandri}
\affiliation{INAF, Osservatorio Astronomico di Roma, Via Frascati 33, I-00040 Monte Porzio Catone (RM), Italy.}

\author[0000-0002-4670-7509]{B. D. Metzger}
\affil{Department of Physics and Columbia Astrophysics Laboratory, Columbia University, Pupin Hall, New York, NY 10027, USA.}
\affil{Center for Computational Astrophysics, Flatiron Institute, 162 5th Ave, New York, NY 10010, USA.} 

\author[0000-0001-9309-7873]{S. R. Oates}
\affiliation{School of Physics and Astronomy \& Institute for Gravitational Wave Astronomy, University of Birmingham, Birmingham B15 2TT, UK.}

\author[0000-0001-8646-4858]{E. Pian}
\affiliation{INAF, Osservatorio di Astrofisica e Scienza dello Spazio, via Piero Gobetti 93/3, 40024, Bologna, Italy.}

\author[0000-0002-8875-5453]{S. Piranomonte}
\affiliation{INAF, Osservatorio Astronomico di Roma, Via Frascati 33, I-00040 Monte Porzio Catone (RM), Italy.}

\author[0000-0003-3457-9375]{G. Pugliese}
\affiliation{Astronomical Institute Anton Pannekoek, University of Amsterdam, 1090 GE Amsterdam, The Netherlands.}

\author[0000-0002-4744-9898]{J. L. Racusin}
\affiliation{Astrophysics Science Division, NASA Goddard Space Flight Center, Greenbelt, MD 20771, USA}

\author[0000-0002-9267-6213]{J. C. Rastinejad}
\affiliation{Center for Interdisciplinary Exploration and Research in Astrophysics (CIERA) and Department of Physics and Astronomy, Northwestern University, Evanston, IL 60208, USA.}

\author[0000-0003-3193-4714]{M. E. Ravasio}
\affiliation{Department of Astrophysics/IMAPP, Radboud University, 6525 AJ Nijmegen, The Netherlands.}
\affiliation{INAF, Osservatorio Astronomico di Brera, Via E. Bianchi 46, I-23807, Merate (LC), Italy.}

\author[0000-0002-8860-6538]{A. Rossi}
\affiliation{INAF, Osservatorio di Astrofisica e Scienza dello Spazio, via Piero Gobetti 93/3, 40024, Bologna, Italy.}

\author[0000-0002-6950-4587]{A. Saccardi}
\affiliation{GEPI, Observatoire de Paris, Universit\'e PSL, CNRS, 5 Place Jules Janssen, 92190 Meudon, France.}

\author[0000-0002-9393-8078]{R. Salvaterra}
\affiliation{INAF, Istituto di Astrofisica Spaziale e Fisica Cosmica, via Alfonso Corti 12, I-20133 Milano, Italy.}

\author[0000-0001-6620-8347]{B. Sbarufatti}
\affiliation{INAF, Osservatorio Astronomico di Brera, Via E. Bianchi 46, I-23807, Merate (LC), Italy.}

\author[0000-0001-5803-2038]{R. L. C. Starling}
\affiliation{School of Physics and Astronomy, University of Leicester, University Road, Leicester, LE1 7RH, United Kingdom.}

\author[0000-0003-3274-6336]{N. R. Tanvir}
\affiliation{School of Physics and Astronomy, University of Leicester, University Road, Leicester, LE1 7RH, United Kingdom.}

\author[0000-0002-7978-7648]{C. C. Th\"one}
\affiliation{Astronomical Institute of the Czech Academy of Sciences, Fri\v{c}ova 298, 251 65 Ond\v{r}ejov, Czech Republic.}

\author{A.J. van der Horst}
\affiliation{Department of Physics, George Washington University, 725 21st St NW, Washington, DC 20052, USA}

\author[0000-0001-9398-4907]{S. D. Vergani}
\affiliation{GEPI, Observatoire de Paris, Universit\'e PSL, CNRS, 5 Place Jules Janssen, 92190 Meudon, France.}

\author[0000-0002-4465-8264]{D. Watson}
\affiliation{The Cosmic DAWN Center.}
\affiliation{Niels Bohr Institute, Copenhagen University, Jagtvej 155, DK-2200, Copenhagen N, Denmark.}

\author[0000-0002-9133-7957]{K. Wiersema}
\affiliation{Physics Department, Lancaster University, Lancaster, LA1 4YB, UK.}

\author{R.A.M.J. Wijers}
\affiliation{Astronomical Institute Anton Pannekoek, University of Amsterdam, 1090 GE Amsterdam, The Netherlands.}

\author[0000-0003-3257-9435]{Dong Xu}
\affiliation{Key Laboratory of Space Astronomy, National Astronomical Observatories, Chinese Academy of Sciences, Beijing, 100101, China.}

\begin{abstract}
We present \jwst\ and {\em Hubble Space Telescope} (\hst) observations of the afterglow of GRB\,221009A, the brightest gamma-ray burst (GRB) ever observed. 
This includes the first mid-IR spectra of any GRB,
obtained with \jwst/NIRSPEC (0.6-5.5 micron) and MIRI (5-12 micron), 12 days after the burst.
Assuming that the intrinsic spectral slope is a single power-law, with $F_{\nu}\propto \nu^{-\beta}$, we obtain $\beta \approx 0.35$, modified by substantial dust extinction with $A_V = 4.9$. 
This suggests extinction above the notional Galactic value, possibly due to patchy extinction within the Milky Way or dust in the GRB host galaxy. 
It further implies that the X-ray and optical/IR regimes are not on the same segment of the synchrotron spectrum of the afterglow. If the cooling break lies between the X-ray and optical/IR, then the temporal decay rates would only match a post jet-break model, with electron index $p<2$, and with the jet expanding into a uniform  ISM medium. The shape of the \jwst\ spectrum is near-identical in the optical/nIR to X-shooter spectroscopy obtained at 0.5 days and to later time observations with \hst. The lack of spectral evolution suggests that any accompanying supernova (SN) is either substantially fainter or bluer than SN~1998bw, the proto-type GRB-SN. Our \hst\ observations also reveal a disc-like host galaxy, viewed close to edge-on, that further complicates the isolation of any supernova component. The host galaxy appears rather typical amongst long-GRB hosts and suggests that the extreme properties of GRB 221009A are not directly tied to its galaxy-scale environment.
\end{abstract}


\section{Introduction} \label{sec:intro}
Gamma-ray bursts (GRBs) are the instantaneously most luminous events known in the Universe. They arise in at least two varieties, long and short \citep{kouveliotou93}, reflecting the typical durations of their prompt emission. The majority of long bursts are thought to arise from the collapse of
very massive stars, an origin secured 
through observations of associated
supernovae \citep{Hjorth+03, levan16}. 
Many of the short GRBs likely arise from
the merger of compact objects, as evidenced
by the presence of kilonova emission in their lightcurves
\citep{tanvir13,berger13,gompertz18,lamb2019,rastinejad21}, and most robustly by their association with
a gravitational wave signal \citep{abbott17a}. Although it should be noted that there is clearly a more significant overlap in the progenitors of long and short-GRBs than previously realized, with supernovae in some short GRBs \citep{ahumada21,rossi22}, and 
kilonovae in bursts with durations in excess of a minute \citep{rastinejad22,troja22,yang22,mei22}. 
GRBs have been used as probes of extreme physics, routes to understanding stellar evolution and as lighthouses to the distant Universe. 

The long-duration GRB\,221009A is, by any measure, the brightest GRB to have been discovered in more than 50 years of sky-monitoring and out of $\sim 10{,}000$ GRBs. Rate estimates suggest bursts like it should occur only once every few centuries \citep{williams2023,malesani23,burns23}. Furthermore, it is the first GRB to have emission detected at tens of TeV \citep{huang22,dzhappuev22}, and its afterglow has been observed from the $\gamma$-ray to radio as part of intensive follow-up \citep[e.g.][]{kann23,laskar2023,oconnor23,williams2023}. Critically, observations from the Very Large Telescope (VLT)/X-shooter \citep{AdUP22,malesani23}, and subsequently the Gran Telescopio Canarias (GTC) \citep{castro-tirado22} showed the redshift to be $z=0.151$ -- a very local event by GRB standards.

Most bursts found at low redshifts have been low-energy events \citep[e.g.][]{soderberg06,chapman07}, perhaps the result of a fundamentally different emission process \citep[e.g., arising from shock break-out rather than directly from the relativistic jet itself][]{campana06,waxman07}. Instead, the isotropic-equivalent energy release of GRB\,221009A is $E_{\gamma,{\rm iso}}>10^{54}$ erg, and comparable to the most energetic and distant GRBs seen at high redshift. Furthermore, the event's proximity is such that any associated supernova, and its underlying host galaxy, are open to intensive study, offering the opportunity to test similarities between the substantially sub-luminous local GRB population and the much more luminous cosmological population. 
However, in the case of GRB 221009A, this is complicated by a location on the sky near the Galactic plane, where foreground extinction is both large 
($A_V \sim 4.2$) 
and uncertain, where crowding complicates optical and IR observations. Even X-ray observations must contend with  
an additional contribution from the dust-scattered 
X-ray halo \citep{williams2023}. 

Despite these challenges, observations to date have yielded a rich data set across the electromagnetic spectrum \citep[e.g.][]{fulton23,laskar2023,williams2023,shrestha2023,kann23,oconnor23}. These data paint a complex picture of a burst with multiple components not readily subsumed within standard afterglow models. In addition, there are apparent detections of the associated supernova SN~2022ixw \citep{fulton23} and excess emission in the radio regime \citep{laskar2023}.

Here we present a set of space-based, high spatial resolution observations of GRB 221009A obtained with \jwst\ and  \hst. These minimize impacts from crowding, extend redward of the limit of ground-based observations and provide the necessary spatial resolution to identify the host galaxy. Although they do not sample the temporal evolution of the event as well as
the extensive observations from the ground, they are, in principle, substantially cleaner because of their high resolution and signal-to-noise ratio (S/N) and provide a well-sampled spectral energy distribution. Therefore, we use them to probe the evolving SED of GRB 221009A, which is essential in understanding both the physics of the blast wave and the presence and properties of any associated supernova. 

\section{Observations}
Many space-based $\gamma$-ray observatories identified GRB 221009A. 
These included {\em Fermi}-GBM \citep{veres22}, {\em Fermi}-LAT \citep{bissaldi22}, {\em AGILE}/MCAL \citep{ursi22}, {\em AGILE}/GRID \citep{piano22}, {\em INTEGRAL} \citep{gotz22}, Konus-{\em Wind} \citep{frederiks22} {\em Insight}-HMXT \citep{tan22}, \textit{STPSat-6}/SIRI-2 \citep{mitchell22}, \textit{SATech-01/GECAM-C} HEBS \citep{liu22}, \textit{SRG}/ART-XC \citep{2022GCN.32663....1L}, {\em Solar Orbiter}/STIX \citep{2022GCN.32661....1X}, 
and \textit{GRBalpha} \citep{ripa22}. The initial brightness seen by the {\em Neil Gehrels Swift Observatory} ({\em Swift}) was sufficiently extreme (and also considering its on sky location
in the plane of the Milky Way) that it was proposed to be
a new Galactic transient rather than a GRB, despite the fact that {\em Swift} triggered on the afterglow emission \citep{2022GCN.32632....1D}

Following the identification of the source as a GRB \citep{2022GCN.32635....1K}, ground-based 
observations rapidly secured a redshift measurement of $z=0.151$ \citep{AdUP22,malesani23}. 
X-ray and optical observations continued until the source entered Sun-block and found a typical GRB afterglow decay. 
The optical data also showed evidence for emission from an accompanying supernova \citep{fulton23},
although, as we will discuss in Section~\ref{SN}, isolation of such a supernova component is challenging. 

\subsection{\em James Webb Space Telescope}
\begin{figure}[t!]
\plotone{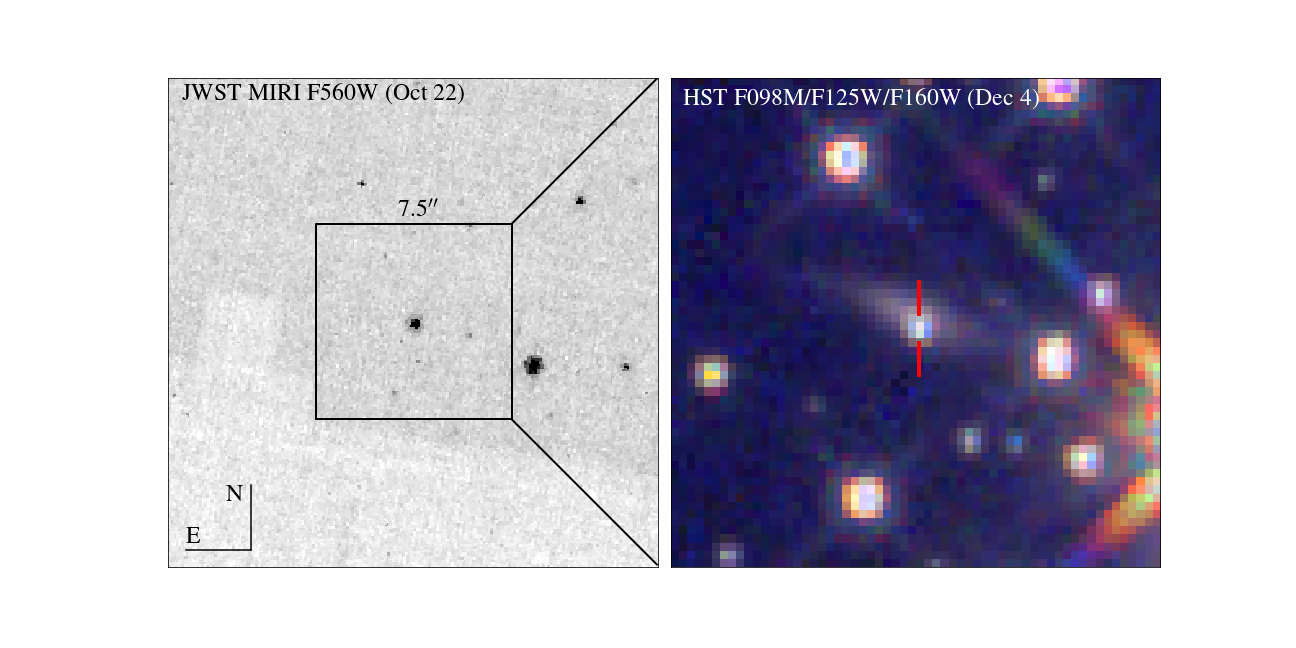}
\caption{The field of GRB 221009A as imaged for target acquisition for the \jwst\ spectroscopy on 21 October (left), and later time observations with \hst\ in the F098M/F125W and F160W filters (right). Only a short sequence was obtained for \jwst, while the \hst\ observations are substantially deeper. These observations clearly show the host galaxy of GRB 221009A extending to the NE to SW of the afterglow position. }
\label{finder}
\end{figure}

\begin{figure}[t!]
\plotone{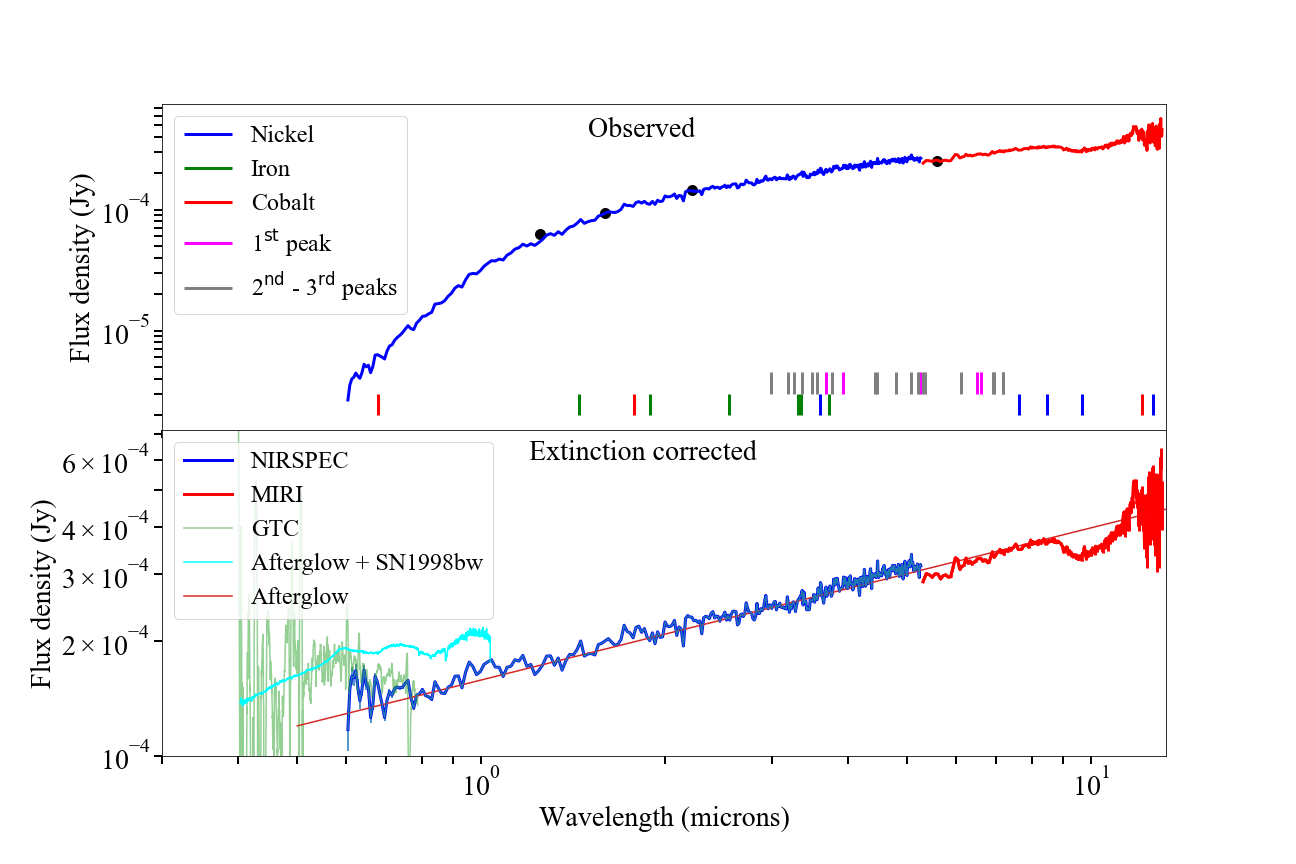}
\caption{The \jwst\ NIRSPEC + MIRI spectrum of GRB 221009A as observed at 12 days post burst. 
The upper panel shows the observed spectrum (with the black points near-simultaneous photometry), while the lower panel is corrected for a foreground extinction of $A_V$ = 4.9. While this has little impact on the
mid-IR, the correction factors for the optical are a factor of $\sim 100$. 
The spectrum appears largely featureless, although some possible absorption features are seen. At this epoch, there are no individual emission features. For 
comparison, expected lines from iron group elements and possible $r-$process contributions are marked \citep{hotokezaka22}.
For comparison, we also plot the best fit absorbed power-law model and how SN~1998bw at 12 days would appear in addition to that model (the SN should contribute minimally at $>5$ microns, such that this additive approach is reasonable). The spectrum does not require the presence of any additional component. However, the bluest regions (which are also those most impacted by the extinction correction) do show an upturn which is apparently also present in GTC observations. This upturn could be indicative of some supernova contribution in the blue. }
\label{lines}
\end{figure}

On 22 October 2022, we obtained observations of the afterglow of GRB 221009A with \jwst\ (programme GO 2782, PI Levan). 
A single, uninterrupted set of observations were obtained with the Near Infrared Spectrograph (NIRSPEC) \citep{jakobsen22} and Mid-Infrared Instrument (MIRI) \citep{rieke15}. 
NIRSPEC observations began at 17:13 UT and MIRI at 18:12, corresponding to times since burst of 
13.16 and 13.20 days, respectively. An image of the field at the time is shown in Figure~\ref{finder}, and the resulting spectra are shown in Figure~\ref{lines}. 

For NIRSPEC, we utilized the prism, spanning a spectral range from 0.5-5.5 microns at a low (and variable) spectral resolution. The MIRI observations were undertaken in low-resolution mode and span the 5-12 micron range. For both NIRSPEC and MIRI observations, we reprocessed the data with the most up-to-date calibrations from December 2022 and obtained 1D extractions. Comparing these products with those obtained from the archive processing shows
good agreement, however, our re-reduction of the NIRSPEC data is $\sim 10$\% brighter beyond 5 microns than the archival data. This reprocessing provides more consistent spectral fits between MIRI and NIRSPEC (see below), although it also introduces a small dis-joint at the overlap region (see Figure~\ref{lines}).

In addition to spectroscopy, a short (11-second) acquisition image was also obtained in F560W with MIRI, as shown in Figure~\ref{finder}. This provides a photometric measurement at this epoch of F560W(AB) = 17.9 $\pm$ 0.1. For 
NIRSPEC data, we used near-simultaneous observations taken with the 3.6\,m Italian Telescopio Nazionale Galileo (TNG) to check the calibration in the J, H and K bands. The match appears excellent, so we adopt the flux calibration direct from the pipeline without additional scaling. 

\subsection{Supporting observations}

In order to build a simultaneous spectral energy distribution at the time of the \jwst\ observations we utilize the {\em Swift} XRT data, as well as obtaining observations at the 10.4\,m GTC in the optical, the 3.6\,m Italian TNG in near-infrared and with the Northern Extended Millimetre Array (NOEMA) in the millimeter regime. 

The GRB was observed with the OSIRIS instrument on the GTC  one day after the \jwst\ observation, 14.31\,days after the burst (program GTCMULTIPLE2M-22B, PI Kann). The observations consisted of both imaging in {\it g}, {\it r}, {\it i}, and {\it z} and spectroscopy. We perform a small correction in the photometry by using the observed temporal decay slope measured from the light curve to derive the photometry at the time of the \jwst\ spectra. The OSIRIS spectroscopy consisted of 4$\times$1200\,s exposures with the R1000B grism, that covers the spectral range between 3700 and 7800\,{\AA} at a resolving power of $R\approx600$.

Near-infrared (NIR) observations of GRB\,221009A were carried out with the TNG telescope
using the NICS instrument in imaging mode. Here we use observations carried out on October 22, approximately 13.3\,days after the burst, and near-simultaneous with the \jwst\ observation.
The image reduction was carried out using the \jitter task of the \ESOeclipse
package\footnote{https://www.eso.org/sci/software/eclipse/}.  Astrometry was performed using the 2MASS\footnote{https://irsa.ipac.caltech.edu/Missions/2mass.html} catalogue. 
Aperture and PSF-matched photometry were performed using the \daophot package \citep{1987PASP...99..191S}. To minimize any systematic effect, we performed differential photometry with respect to a selection of local isolated and non-saturated reference stars from the 2MASS and the UKIDSS\footnote{http://www.ukidss.org/} surveys. The resulting magnitudes in the AB-system are J=19.41 $\pm$ 0.06, H=18.97 $\pm$ 0.06 and K = 18.49 $\pm$ 0.08 at times of 13.25\,days post burst.

A set of millimeter data taken with NOEMA between 78 and 150 GHz (program S22BF, PI de Ugarte Postigo) were interpolated to the epoch of the \jwst\ observation. The observations were performed in the medium-extended C configuration. The data reduction and analysis was done with \clic and \mapping from the \gildas software package, flux calibration was relative to the reference sources MWC349 and LKHA101. The fluxes were determined with UV point-source fits for a consistent error propagation. A full analysis of the NOEMA dataset will be published in a forthcoming paper.

In addition to data taken near simultaneously with \jwst\, we also utilized the observations with VLT/X-shooter obtained at 0.5 days post-burst \citep{malesani23} since these provide an ideal comparison epoch where the source should be dominated by purely afterglow emission (i.e. the epoch is so early and the afterglow so bright that there should be no supernova emission). The data reduction is described in detail in \cite{malesani23}.

\subsection{Hubble Space Telescope}
We obtained three epochs of imaging 
with the {\em Hubble Space Telescope} (\hst) on 8 November 2022, 19 November 2022 and 4 December 2022 (Program 17264, PI: Levan), corresponding to $\sim 30, 41$ and 56 days post burst. At the first epoch, observations were obtained in F625W, F775W, F098M, F125W and F160W. A guide star failure in the IR observations during the second epoch meant that observations were only obtained in the optical. At the final observations on 4 December, we obtained F625W, F098M, F125W and F160W. The data were aligned and reduced via {\tt astrodrizzle}, while the native pixel scale was retained due to the relative paucity of dithers in most cases (0.04 arcsec/pix for WFC3/UVIS images and 0.13 arcsec/pix for WFC3/IR images). 

The images clearly show the afterglow in all bands superimposed on an underlying host galaxy. This host galaxy contributes modestly at the time of the first epoch but more than 50\% of the light in $1^{\prime\prime}$ apertures at later times. Ultimately, the optimal way to remove the host contribution would be via the direct subtraction of late-time images. However, given the brightness of GRB 221009A, the afterglow contribution may remain detectable to \hst\ for several years. We, therefore, report photometry via three different approaches. 

\begin{figure}[t!]
\plotone{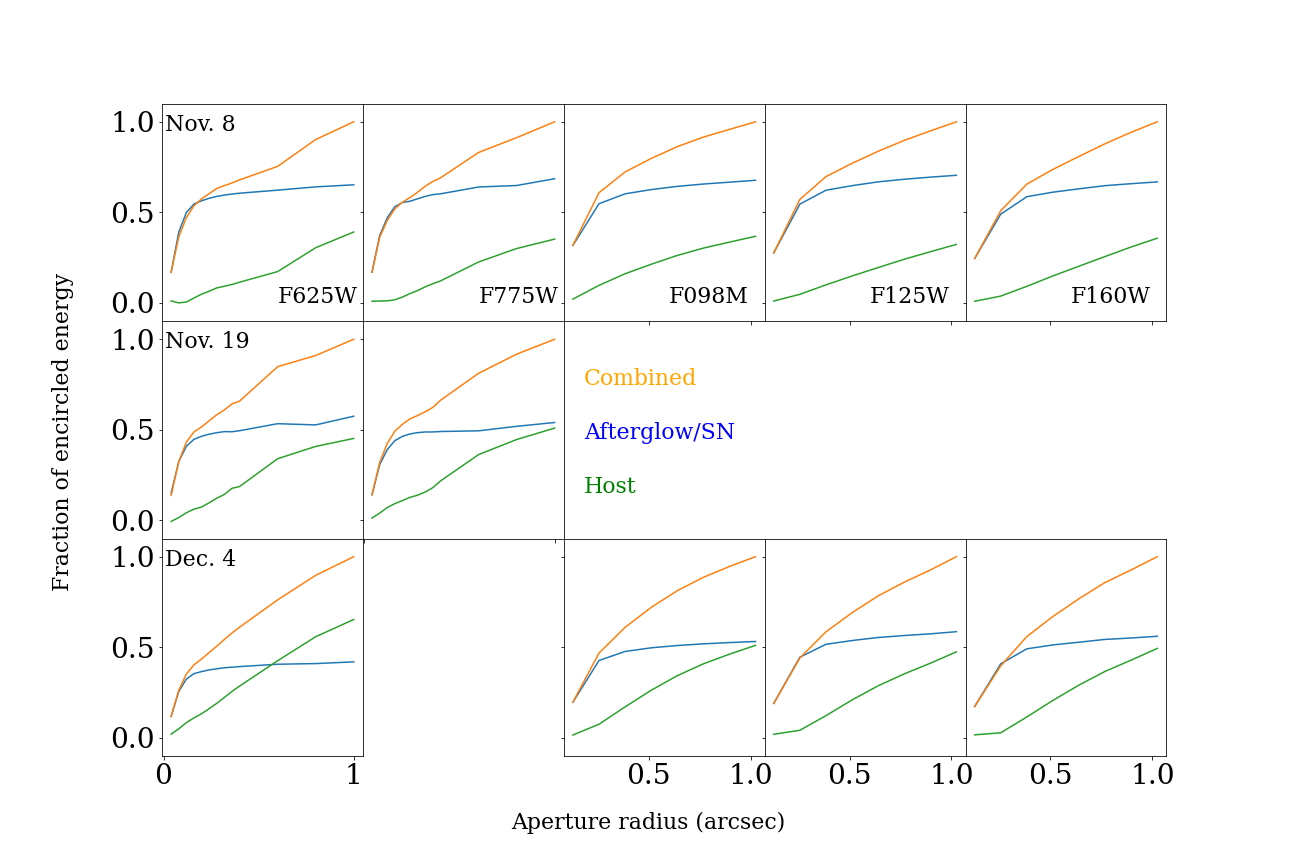}
\caption{Radial profiles of the afterglow and host of GRB 221009A and their decomposition. The orange line shows the observed radial profile. The blue line is a stellar source from the field, scaled to the central regions such that the residual in the central $\sim 0.1^{\prime\prime}$ is
comparable to the flux value in a region of the host at a similar offset to the GRB. The green line shows the residual following the subtraction of that point source, and represents the host galaxy light. The panels are normalised to the total enclosed counts within a $1^{\prime\prime}$ 
aperture around the afterglow position.} 
\label{decomposition}
\end{figure}

Firstly, we measure the curve of growth around the afterglow location, compared to that of an isolated star within the image (see Figure~\ref{decomposition}). We then scale the point spread function and subtract it. In undertaking this subtraction, we consider the case where all the light at the afterglow position is provided by the afterglow (i.e. we subtract to zero) and where there is an underlying contribution from the host galaxy. We estimate this contribution by measuring the flux in a small aperture at a location on the host a comparable distance from the centre as the afterglow. We assign this value as our best estimate of the actual afterglow flux and set the error as the difference between this and zero (when applied symmetrically, this allows for either no underlying host contribution or a relatively bright underlying star-forming region). We believe this provides a conservative error estimate for the actual afterglow/SN brightness at any given epoch and note that this error is substantially larger than the photon counting/background errors introduced via the photometry (which are typically $<1$\%). 

Secondly, we also report measurements made in a larger ($1^{\prime\prime}$) aperture. This is the largest aperture that can be used without introducing significant additional light from other sources in the crowded field of view. These magnitudes are comparable to those measured from the ground and may be helpful for ground-based comparisons. 

Finally, the unambiguous detection of the host galaxy in the F098M, F125W, and F160W filters at the last \hst\ epoch (Figure~\ref{finder}) provides a different approach to decomposing the emission through modelling the light profile of the host. To model the system, we use a parametric method based on \galfit \citep{Peng2010} to simultaneously fit a \sersic function for the host and a point spread function (PSF) component for the afterglow \citep[see][for more details on the method]{Schneider2022}. In addition, we also exploit the available multi-wavelength images of the host to perform a multi-band fit and derive a more robust wavelength-dependent model. This model is derived using \galfitm \citep{Haussler2013,Vika2013}, an extended multi-band version of \galfit where each component parameter is replaced by a polynomial function of wavelength. \galfitm is expected to provide a more consistent and homogeneous model of the object over wavelength and improve the information extracted from lower signal-to-noise (S/N) ratio bands.

First, we ran \galfitm to simultaneously model the infrared filters (F098M, F125W, F160W) of the final observation epoch (4 December). For the \sersic function parameters, we consider a common position ($x_c$, $y_c$), axis ratio $(b/a)$ and position angle (PA) for the three filters and let them vary as a constant offset from the input value. The half-light radius (also known as the effective radius and denoted as \re) and \sersic index ($n$) are defined as a linear function of the wavelength, while the magnitude is defined as a completely free parameter. Similarly, we use a PSF model with a free magnitude and a constant position as a function of the wavelength for the afterglow. Relaxing these assumptions does not strongly affect our estimates, especially for the magnitudes, $R_{\rm e}$, and \sersic index. 

\begin{figure}
     \centering
     \includegraphics[width=0.49\textwidth]{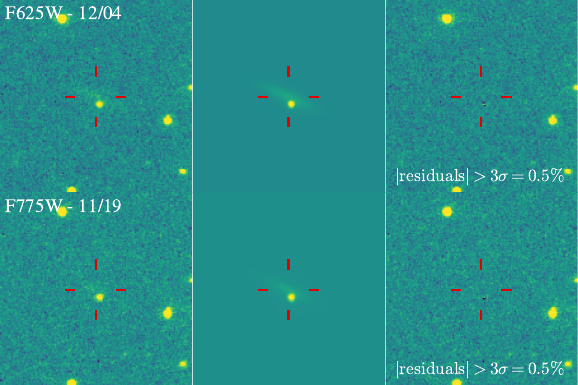}
     \includegraphics[width=0.49\textwidth]{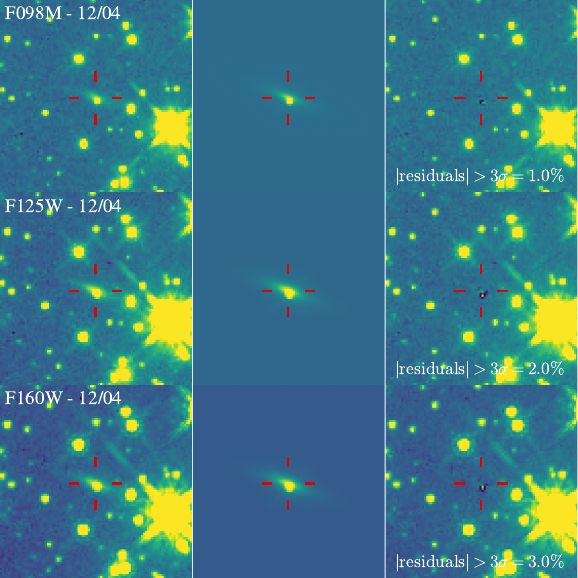}
     \caption{The five \hst\ data sets used to decompose the host and afterglow light. Each panel row corresponds to a different filter with \hst\ images of the GRB~221009A field on the left, the best-fit \galfitm models on the middle, and the residual maps on the right. Panels are centered on best-fit host positions and oriented north to the up and east to the left. A common color scale is considered for a given row. The UVIS (left part) and IR (right part) filters correspond to a square region of 
     $6^{\prime\prime}$ 
     and 
     $12^{\prime\prime}$, 
     respectively. The fraction of pixels with a residual larger than $3\sigma$ within a constant 
     $1^{\prime\prime}$ 
     aperture radius is shown in the lower part of the residual map.}
     \label{fig:galfit}
\end{figure}

We adopted a similar approach to fit the \hst/UVIS filters simultaneously. However, only F625W observations were secured during the last \hst\ observations. We instead consider the 19 November observation for the F775W filter. Given that only the afterglow magnitude is expected to vary significantly between the two epochs, the observational delay on the host galaxy model should be limited. It is worth noting that the exposure time of F625W is about five times longer than F775W exposure and thus provides a deeper and higher S/N ratio image to drive and model the host through \galfitm. As a sanity check, we also run \galfit on each filter individually. We find consistent host models with those of \galfitm, except for F775W, which converges to a different solution. This is likely due to the higher Galactic extinction that affects the host emission at this wavelength. The best-fit models and residual maps determined by \galfitm for the five \hst\ filters are shown in Figure~\ref{fig:galfit}. A visual inspection of the residual maps confirms that we successfully reduced the majority of the initial flux of the system. More quantitatively, we consider a constant $1^{\prime\prime}$
radius aperture at the host position to measure the fraction of pixels above $3\sigma$ before and after the subtraction. We find that this fraction is reduced by more than 90\% for all bands. We also note the presence of a marginal over-subtracted signal for the IR filters at the afterglow position. This might be caused by the central core of the PSF model used for the fit or by the presence of a compact and unresolved active star-forming region at the burst location, frequently observed for long GRB host galaxies \citep[e.g.,][]{2006Natur.441..463F,lyman17}. 

Once the best-fit model of the host galaxy was determined, we used it as a constant input for all \hst\ images. We thus run \galfit with a constant\footnote{More precisely, we let vary the host component within the uncertainties of the best-fit model.} \sersic model for the host plus a free PSF model for the burst. The afterglow magnitudes derived from this approach are reported in Table.~\ref{phot}, and we believe these to be the most robust estimates of the afterglow magnitudes at these epochs. The structural parameters of the host are further discussed in Section~\ref{sec:host_galaxy}.

\section{Spectral shape and evolution} 

\subsection{Optical to mid-IR spectral shape} 
We first consider the spectral shape observed in the NIRSPEC and MIRI data. The 
calibration appears robust, with the NIRSPEC observations matching ground-based JHK observations. The normalisation of the MIRI data is consistent with the F560W acquisition image. 

The spectrum is highly absorbed due to a significant Galactic foreground. This manifests as a strong
suppression of the optical flux, while silicate features are also visible in the MIRI band at 8-10 microns. The strength of 
these silicate features likely varies on different lines of sight, and straightforward extinction laws do not remove it. Therefore the region of the spectrum between $\sim 3-8$ microns is likely to be least affected by Galactic extinction and can 
provide an estimate of the spectral slope. Indeed, this appears to be blue with 
$\beta \sim 0.4$ (defined as a power-law
with $F_{\nu} \propto \nu^{-\beta}$). This is much bluer than the X-ray spectral slope $\beta = 0.91 \pm 0.09$ at approximately the same time \citep{williams2023}. 
This would be consistent with the presence of the cooling break between the two bands, although the very blue spectral slope is not naturally expected. 

We perform a joint fit to the NIRSPEC and MIRI observations to better quantify the spectrum.
We perform fits for two models:
(a) a power law plus Milky Way (MW) extinction;
and (b) a power-law plus host galaxy and Milky Way extinction.
We set priors on the model parameters of $0.0 \leq \beta \leq 1.0$ for the power-law spectral index. 
We allow for a wide range of extinction within the MW ($0.0 \leq A_{V,{\rm  MW}} \leq 10.0$) and allow the fit to identify the best value given the condition $A_V \geq R_V E(B-V)$.
We take the $A_V/E(B-V) = 2.742$ from Table 6 in \cite{schlafly11}, and using an initial $A_V = 4.17$ we find $E(B-V) = 1.52$. This value is used to set the maximum possible $R_V$ for a given $A_V$ such that $0.0 < R_V \leq A_V/E(B-V)$.
In case (b), the prior on the host extinction is drawn from an exponential distribution, 
$P(A_{V,{\rm  host}}) \propto \exp[-A_{V,{\rm  host}}/0.3]$ \citep[e.g.][]{holwerda2015}, with a fixed $R_V=2.93$, and the MW extinction is the same as that in case (a). 

\begin{figure}
     \centering
     \includegraphics[width=\textwidth]{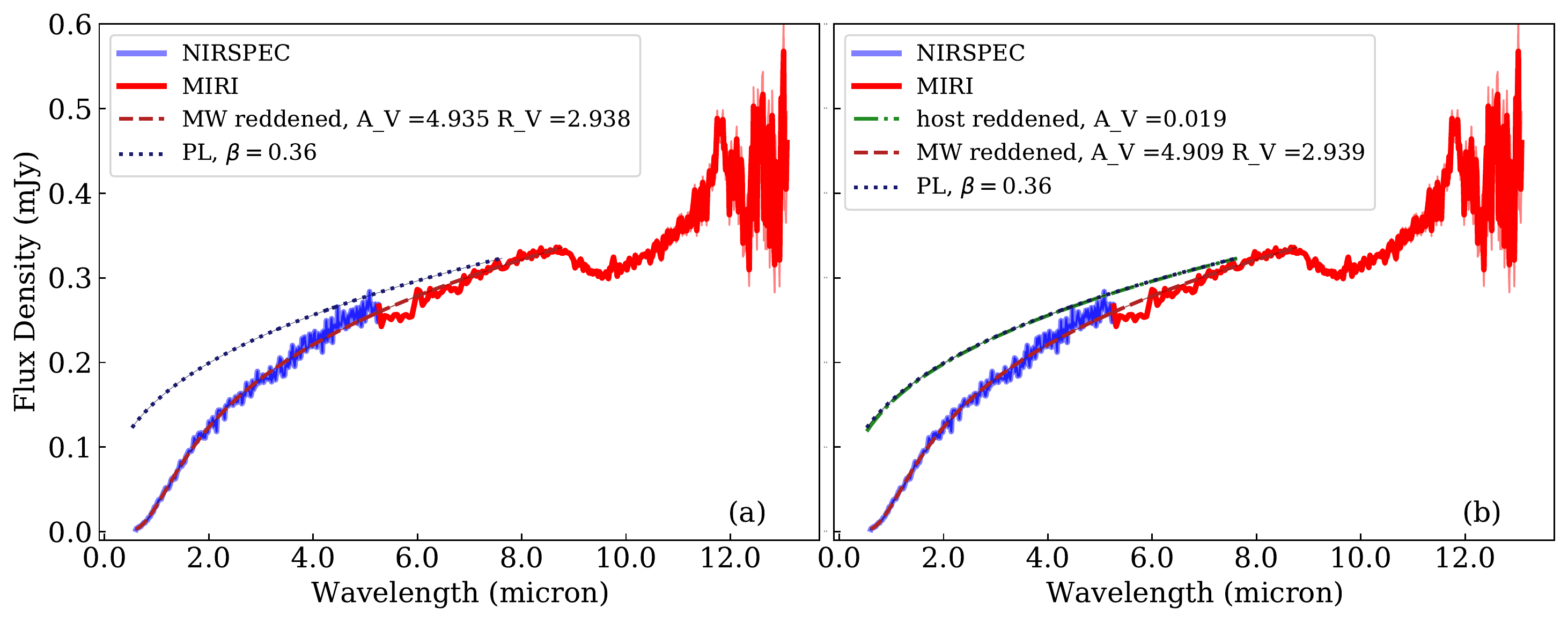}
     \caption{Fits found via MCMC for the two models: (a) a power-law plus Milky Way extinction; and (b) a power-law plus host galaxy and Milky Way extinction.}
     \label{fig:spec_fits_jwst}
\end{figure}

For both the Milky Way and the host galaxy, we use a \citet{fitzpatrick99} extinction curve at $z=0.0$ and $z=0.151$, respectively.
The models are fitted to the observed spectrum at wavelengths $<8.8$\, micron, to avoid any bias in the fits due to a more complex extinction feature at longer wavelengths. 
Figure \ref{fig:spec_fits_jwst}, plotted against the joint data, shows our resultant spectral fits.
Both models give a good approximation to the observed spectral shape, and the total extinction is consistent between them.
The posterior distribution for model b) shows a tight correlation between the Milky Way and host extinction, indicative of a degeneracy between these parameters.
The extinction parameter and spectral index values from the Markov-Chain Monte Carlo (MCMC) are shown in Table \ref{tab:post}.

We note that the NIRSPEC and MIRI spectra have a statistically significant but modest difference in the preferred power-law index.
By fitting a broken power law model to the joint spectrum for NIRSPEC and MIRI, at $<8.8$\,micron, we find a best-fit model, using a reasonably sharp transition in a smoothly broken power law.
The break is found at $\lambda \sim 4.46$\,micron, and with a power law index below the break i.e., at shorter wavelengths, of $\beta = 0.40$ and above the break, longer, of $\beta=0.32$.
The sharpness of this break, its near coincidence with the point where the two spectra join, and the short wavelength range for the fit, especially with the MIRI data, suggests this difference, although statistically significant, is unlikely to be related to a physical change in the source spectrum. We, therefore, use the single power-law results.

\begin{figure}[t!]
\plotone{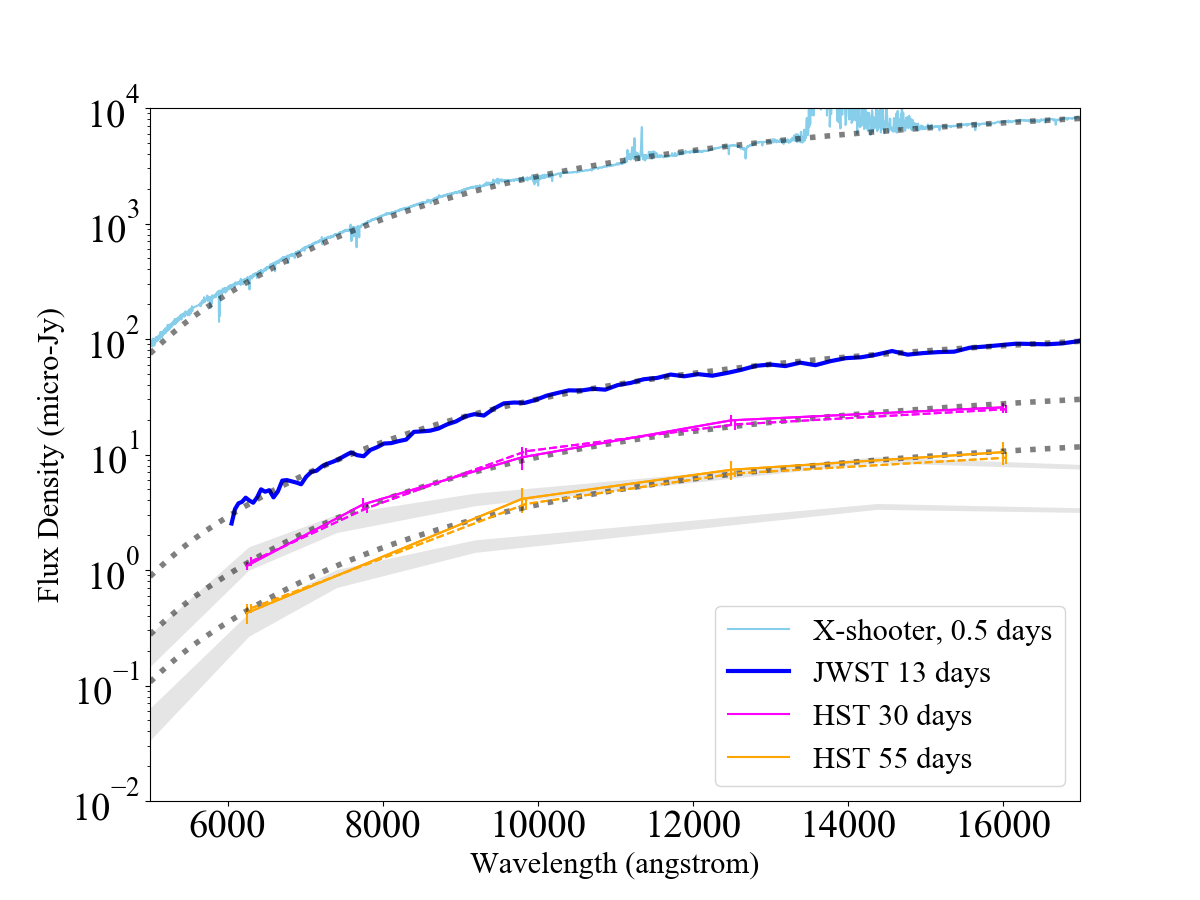}
\caption{The evolution of the spectral energy distribution of the GRB 221009A afterglow from 0.5 to 55 days. The figure shows a 0.5-day X-shooter spectrum, the 12-day \jwst\ spectrum, and \hst\ observations at 30 and 55 days. The solid and dashed lines refer to the decomposition of \galfit\ estimates of the point source magnitude and have been slightly offset in wavelength for clarity. The dashed line shows our best fit absorbed power-law model (for a single Galactic extinction fit to our \jwst\ spectrum). As can be seen, there is no evidence for spectral evolution between 0.5 and 12 days, indicating no significant supernova contribution at this time. At later times, the uncertainty in the underlying host galaxy light dominates the \hst\ observations and precludes drawing firm conclusions. However, it also does not support the presence of a bright supernova component. The shaded grey areas show the spectral energy distribution of SN~1998bw at comparable epochs, representing a range of possible extinction from $A_V =4.2-4.9$ mag. 
The dotted lines show the shape of the afterglow as determined from our \jwst\ observations, but scaled to the other epochs. They demonstrate the possible changes in spectral shape from 0.5 - 55 days}.
\label{SED}
\end{figure}

We also fit identical models to the X-shooter spectrum obtained at $\sim 0.5$ days post burst and presented in \cite{malesani23}. This spectrum is dominated entirely by the afterglow (no contribution from either a supernova or host galaxy). It yields $\beta=0.207$, $A_V =4.903$ and $R_V =3.225$, broadly comparable to the extinction values obtained from the \jwst\ observations, although with a notably bluer spectral slope. However, we also note that, observationally, the spectra appear extremely similar (see Figure~\ref{SED}), and the differing values may reflect the lack of redder coverage for the X-shooter observations.

\begin{table}
    \caption{MCMC posterior parameter values for models a), and b), see text, fit to the \jwst/NIRSPEC and MIRI spectrum at $\sim$13 days. Columns are: the source power-law spectral index, $\beta$, the Milky Way extinction, $A_{V,{\rm MW}}$, and the host galaxy extinction, $A_{V,{\rm host}}$.}
    \centering
    \begin{tabular}{c|c|c|c|c}
         Model & $\beta_1$ & $R_V$ & $A_{V,{\rm MW}}$ & $A_{V,{\rm host}}$ \\
         \hline
         a) & $0.362^{+0.001}_{-0.001}$ & $2.938^{+0.008}_{-0.008}$ & $4.935^{+0.006}_{-0.006}$ & -- \\
         b) & $0.362^{+0.001}_{-0.001}$ & $2.939^{+0.008}_{-0.008}$ & $4.909^{+0.019}_{-0.040}$ & $0.019^{+0.030}_{-0.014}$ 
    \end{tabular}
    \label{tab:post}
\end{table}

By introducing a blackbody component with a limited temperature range and a broad luminosity distribution, i.e., consistent with the expectation from a supernova, we can find a limit on the potential supernova contribution to this spectrum. 
The fits return a strong correlation between maximum luminosity and the blackbody temperature, where higher temperatures allow a higher maximum luminosity with the minimum luminosity defined by the prior.
The fits that have a SN-like blackbody temperature, $T_{\rm eff}\gtrsim 5000$\,K return luminosities around $<10^{42}$ erg s$^{-1}$ \citep[approximately 10-20\% of SN~1998bw at the same epoch,][]{nakamura01}. 
We note that the presence of a component with luminosity similar to SN~1998bw
should result in a significant enhancement in the
optical regime (see Figure~\ref{lines}).
Therefore, the best model to describe the \jwst\ spectrum of GRB 221009A has no measurable contribution from a thermal, supernova-like component. However, we cannot rule out the possibility of a very blue supernova, to which our observations have minimal sensitivity.

\subsection{A simultaneous multi-wavelength SED} 
\begin{figure}
     \centering
     \includegraphics[width=\textwidth]{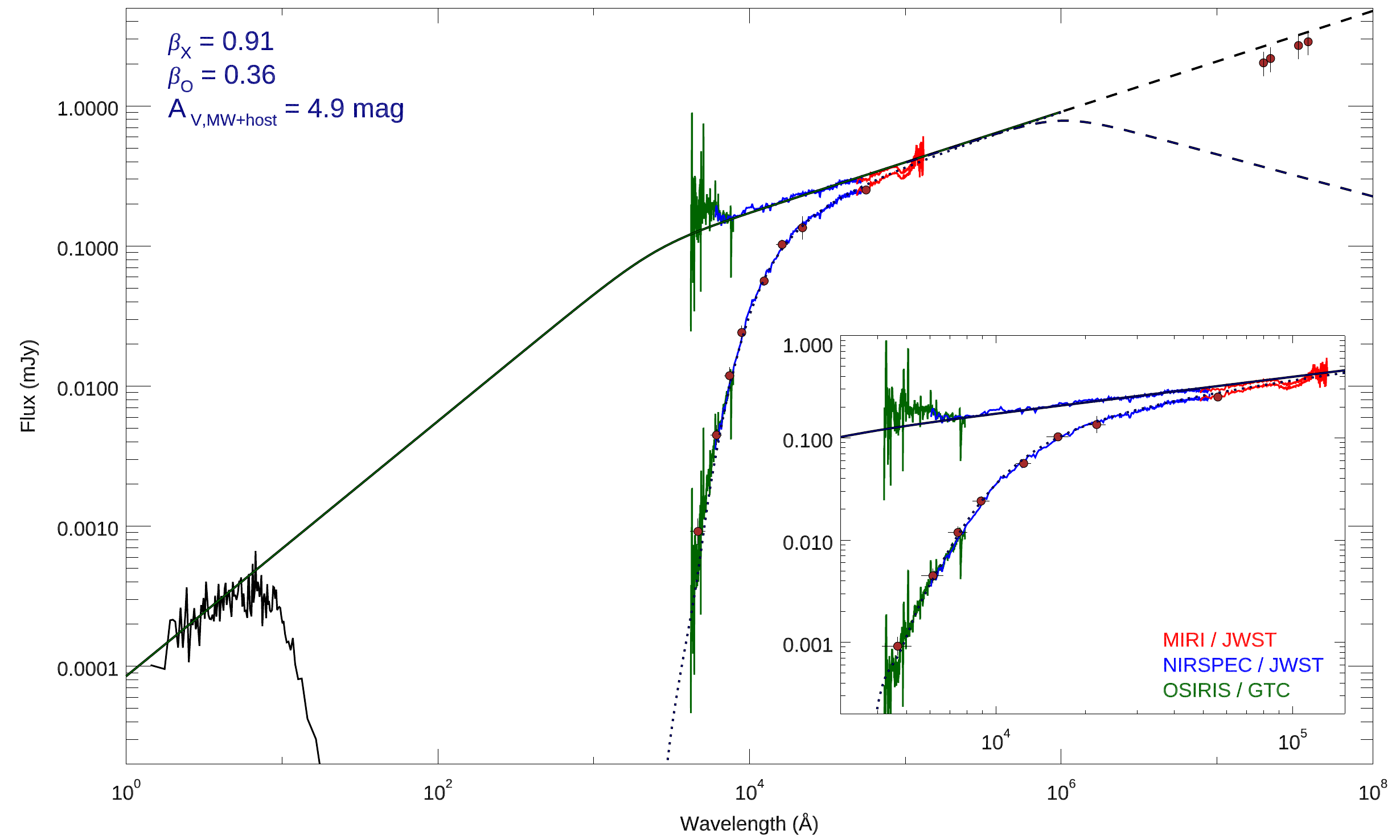}
     \caption{The X-ray ({\em Swift-XRT} to radio (NOEMA) SED of GRB 221009A at the time of the \jwst\ observations.  Solid lines show the underlying model, while the dotted lines are the observed points, with the strong deviation due to the heavy foreground extinction. Redward of the \jwst\ observations we show two possible models. One in which the peak frequency lies just beyond the \jwst\ band \citep[e.g.][]{laskar2023}, and one which extends the \jwst\ spectral slope out towards the radio regime.} Interestingly, the continuation of the \jwst\ spectrum is close to the millimeter points, as expected from our favoured afterglow scenario (see \S \ref{sec:afterglow}).
     \label{fig:spec_fits}
\end{figure}

In addition to our \jwst\ observations, we also build a near-simultaneous broader band spectral energy distribution using X-ray data from the {\em Swift}-XRT \citep[see also][]{williams2023}, our GTC spectroscopy and NOEMA millimeter data. 
Extrapolating the X-ray flux and spectral slope confirms that a break is required between the X-ray and optical/IR regime. The difference in the measured spectral slopes is $\Delta \beta \sim 0.5$, consistent with interpreting this break as the cooling break. This model is shown in Figure~\ref{fig:spec_fits}.   

 The millimetre photometric points appear broadly consistent with the extrapolation of the $\beta=0.35$ slope to this regime. However, a single component from the optical to millimeter is disfavoured due to the different temporal behaviour of these regimes  \citep[see][]{laskar2023}.

\begin{table}
\caption{\hst\ observations of GRB 221009A}
\begin{center}
\begin{tabular}{llllllll}
\hline
Date & MJD & $\Delta$ t (d) &  Band & exp.(s) & Point source & 1$^{\prime \prime}$ aperture & GALFIT \\
\hline
2022-11-08:06:32:47 &59891.27277367 &  29.72 & F625W &  960 & 23.81 $\pm$ 0.08 &  23.13 $\pm$ 0.07 & 23.72 $\pm$ 0.09 \\
2022-11-08:06:41:33  &59891.27886145  & 29.73 &  F775W & 750 & 22.48 $\pm$ 0.13 & 21.97 $\pm$ 0.05 & 22.56 $\pm$ 0.09 \\
2022-11-08:08:08:26  &59891.33919124 & 29.79 & F125W & 798  & 20.66 $\pm$ 0.11 & 20.26 $\pm$ 0.02 & 20.75 $\pm$ 0.11 \\
2022-11-08:08:15:37 &59891.34417976  & 29.79 & F098M  & 898 & 21.46 $\pm$ 0.22 & 20.91 $\pm$ 0.02 & 21.33 $\pm$ 0.07  \\
2022-11-08:08:23:41 & 59891.34978161 & 29.80& F160W  & 798 & 20.38 $\pm$ 0.10 & 19.87 $\pm$  0.01 & 20.42 $\pm$ 0.08 \\
2022-11-19:04:58:39 & 59902.18082923 & 40.63  & F625W & 960 & 24.14 $\pm$ 0.11 & 23.44 $\pm$ 0.07 & 24.20 $\pm$ 0.16  \\
2022-11-19:05:06:15 & 59902.18691700& 40.63 & F775W & 750 & 23.15 $\pm$ 0.23 &  22.32 $\pm$ 0.12 & 23.04 $\pm$ 0.16 \\
2022-12-04:02:02:24 & 59917.05554104 & 55.50 & F625W & 3776 & 24.83 $\pm$ 0.20  &  23.75 $\pm$ 0.10 & 24.73 $\pm$ 0.08 \\ 
2022-12-04:22:03:49 & 59917.84907676&  56.30& F125W & 698 &21.73 $\pm$ 0.25  & 20.91 $\pm$ 0.02 & 21.81 $\pm$ 0.06 \\
2022-12-04:22:11:50 & 59917.85348639 & 56.30 & F098M & 898 & 22.36 $\pm$ 0.19 & 21.53 $\pm$ 0.04 & 22.47 $\pm$ 0.12  \\
2022-12-04:22:18:14 & 59917.85908824 & 56.31 & F160W & 698 & 21.35 $\pm$ 0.23 & 20.49 $\pm$ 0.02 & 21.47 $\pm$ 0.08  \\
\hline
\end{tabular}
\end{center}
\tablecomments{Photometry of the counterpart of GRB 221009A as observed with the {\em Hubble Space Telescope}. The different magnitude columns are for the afterglow magnitude as derived via decomposition of a point and extended source, those measured in a large aperture, and the point source magnitudes from GALFIT.}
\label{phot}
\end{table}

\subsection{Line features} 

In addition to mapping the overall afterglow, we may also expect to observe broad spectral features in these observations related to either the expected underlying supernova emission or the presence of $r-$band nucleosynthesis, which is suggested to occur in the accretion discs formed during long GRBs \citep{siegel19,barnes22}. In Figure~\ref{lines}, we plot the \jwst\ spectrum and mark the locations of prominent iron group lines seen in supernovae, in addition to suggested $r$-process lines from
\cite{hotokezaka22}. There are no apparent lines visible in 
the spectrum. This can be understood as a consequence of the early time of the observations. Firstly, at this epoch, the supernova (if present) is likely not optically thin, and heavier element lines (e.g. in particular, those from the $r$-process) may only be visible in the case of substantial mixing. Secondly, most (if not all) of the light observed via \jwst\ is from the afterglow, not the supernova component.

We note the presence of an apparent feature at $\sim 11.8$ microns in the MIRI band. This would be consistent with a blue-shifted [Co II] line -- a line which was strikingly the strongest observed feature in the mid-IR spectrum of SN~1987A \citep{aitkin88}. 
However, an inspection of the 2D MIRI images, before spectral extraction, suggests a background defect of unknown origin that is broader than the trace, and complicates spectral extraction at this wavelength.
We therefore believe this feature is most likely spurious.

\subsubsection{Supernova emission} 
\label{SN}
As a luminous long-duration GRB, 
we expect emission from an associated supernova explosion to rise in brightness in the days following the burst.
Indeed, both spectroscopic \citep{GCN32800} and
photometric \citep{GCN32818,fulton23}, observations have claimed the detection of the associated supernova, named SN 2022ixw. However, in contrast to these works, we do not see significant evidence for supernova emission in our observations. While we lack the 
temporal resolution of other observations, we do have better spectral coverage, and our spectral energy distribution shows little change from 0.5 to 55 days (Figure~\ref{lines} and Figure~\ref{SED}).

At later times, there also appears to be minimal spectral change in our \hst\ observations (see Figure~\ref{SED}). Although our results between different methods (e.g. \galfit vs curve of growth) are consistent, there are minor differences in the resulting photometry. It does appear that the F160W points lie slightly below the pure afterglow extrapolation in all cases, while the F098M points lie slightly above it. This would be consistent with the presence of some supernova light which should peak around the Y-band with the heavy extinction. However, there is little ability to add an optical-peaking SN component without violating the observed r-band observations. 

To check this, we initially fit a simple scaling of the best fit absorbed power-law (from the \jwst\ data) to the \hst\ photometry obtained via 1-D decomposition or \galfit. These give $\chi^2$/dof for each epoch of 1.1 and 0.7 for the 1-D decomposition and 2.2 and 1.1 for the \galfit\ models. In 3/4 cases, there is little justification for the addition of a further component, although the early (30-day) \hst\ epoch is a relatively poor fit for the \galfit\ values. To quantify the possible contribution of an SN similar to SN~1998bw, we then fit a linear combination of an SN~1998bw template (via the lightcurves of \cite{clocchiatti11}, supplemented with the IR observations of \cite{patat01}). These suggest that an SN between 10-40\% of the brightness of SN~1998bw would improve the fit. However, the presence and properties of the supernova remain subject to significant systematic uncertainty due to the contribution of the underlying host galaxy, which complicates precision photometry. 

The light curve in the \hst\ observations is consistent with a single power-law of slope $\alpha \approx -1.5$ (although with limited 
coverage to distinguish any variation from a power-law). Notably, this slope is consistent with optical/IR measurements made earlier in the afterglow phase. 
Ultimately, \hst\ observations should provide extremely high S/N ratio measurements of the afterglow/supernova brightness. Indeed, given some evidence for late time energy injection in the X-ray \citep{williams2023}, measuring changes in the spectral shape and not just the temporal decay is particularly important. For this, the host galaxy
must be accurately removed, requiring late-time observations. However, for a $t^{-1.4}$ decay for the afterglow, and on the assumption that the IR was dominated by afterglow emission at 55 days post burst, it will take a decade for the afterglow to reach F160W$>$28, where we can be confident of little contribution from the supernova/afterglow. It may be possible to use the differing decays of afterglow and supernova (e.g. a power-law vs an exponential decay from $^{56}$Ni) to decompose lightcurve contributions at earlier epochs, but doing so will require multiple further epochs of \hst\ observations. 

The \jwst\ spectrum should also provide strong constraints on the presence of any supernova. In particular, it is reasonably described by a single power-law without requiring additional broad features from the associated supernovae. Supernovae peak at around the rest-frame V-band, but the combination of redshift and, in particular, the heavy extinction pushes this peak into the near-IR for GRB 221009A \citep[see e.g.][]{fulton23}. In Figure~\ref{lines}, we plot the best-fit power-law to the data after correction for the extinction. The \jwst\ spectrum is in excess of the model at $\sim 1$ micron and appears to rise at the blue end. However, we caution that this region is where the afterglow is faintest and where the extinction correction is largest. However, the simultaneous GTC spectrum also shows rising flux blueward of the \jwst. This may suggest that the blue upturn in \jwst\ is accurate and has a blue supernova contribution. This supernova would be substantially bluer than SN~1998bw at the same epoch. Still, there is some variation in the apparent colors of SNe associated with GRBs. For example, SN 2013cq/GRB 130427A is apparently bluer \citep{levan13,melandri13} than SN~1998bw in the optical (e.g. 4000-8000\,\AA) regime. This possibility will be investigated further by de Ugarte Postigo et al. (in prep). 

Possible explanations for the differing interpretations of the supernova in GRB 221009A may arise from assumptions in the afterglow model. For example, \cite{fulton23} assume that the X-ray and optical lie on the same branch of the power-law spectrum (and hence decay at the same rate). The different spectral slope inferred from the \jwst\ observations suggests this is not the case, and so the assumption of similar decay rates may not be correct. Indeed, fitted separately, the X-ray and optical regimes give different decays (see section~\ref{afterglow_disc}). Indeed, \cite{williams2023} also find a break necessary, although place this break within the X-ray band. It is also possible that a rising supernova component compensates for a spectral change in the afterglow. However, such an explanation would suffer from a fine-tuning problem.

\section{Host galaxy} 
\label{sec:host_galaxy}
The host galaxy of GRB 221009A is visible in our late time \hst\ observations. It appears to be an edge-on system with the burst close to the nuclear regions, but with a notable offset ($0.25^{\prime\prime}$ $\approx$ 0.65~kpc) from the nucleus of the host. At the time of our latest observations, there remains a substantial afterglow contribution. However, our fits with \galfitm allow us to extract reasonably robust host galaxy photometry in the case that a smooth distribution represents the entire galaxy. It is also possible (and perhaps even likely) that the distribution is not smooth due to a bright star-forming region under the GRB position. However, we report here the results based on a single \sersic component. 
Our model magnitudes are F625W~=~24.88~$\pm$~0.08, F775W~=~23.80~$\pm$~0.14, F098M~=~22.00~$\pm$~0.06, F125W~=~21.37~$\pm$~0.07, F160W~=~20.92~$\pm$~0.10. 

Corrected for the tabulated foreground extinction 
provides F625W~=~21.4 or M$_{F625W} \sim$\,-18.0. 
This absolute magnitude is rather typical for a long GRB host galaxy.  
In principle, the colors of the host galaxy can provide details of the stellar population within the host. However, in this case, it is complicated by the large and uncertain foreground extinction. 
We note that the observed colours of the source for this foreground are reasonable, with F625W$-$F160W(AB)~$\approx$~1.2 after extinction correction for the tabulated Milky Way extinction, 
consistent with typical colours for GRB host galaxies as a whole \citep{hjorth12,lyman17}.

\begin{figure}
     \centering
     \includegraphics[width=0.49\textwidth]{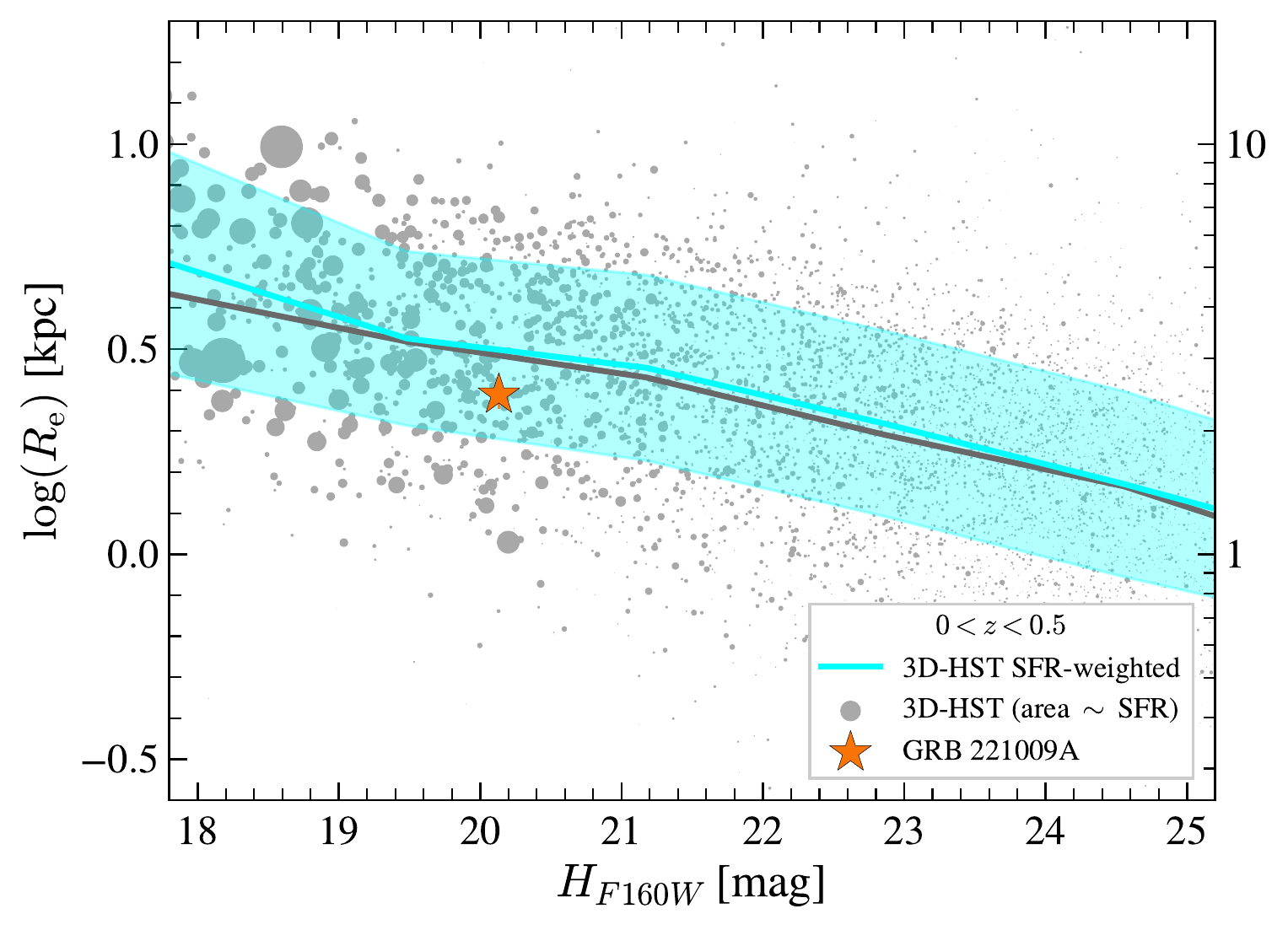}
     \includegraphics[width=0.49\textwidth]{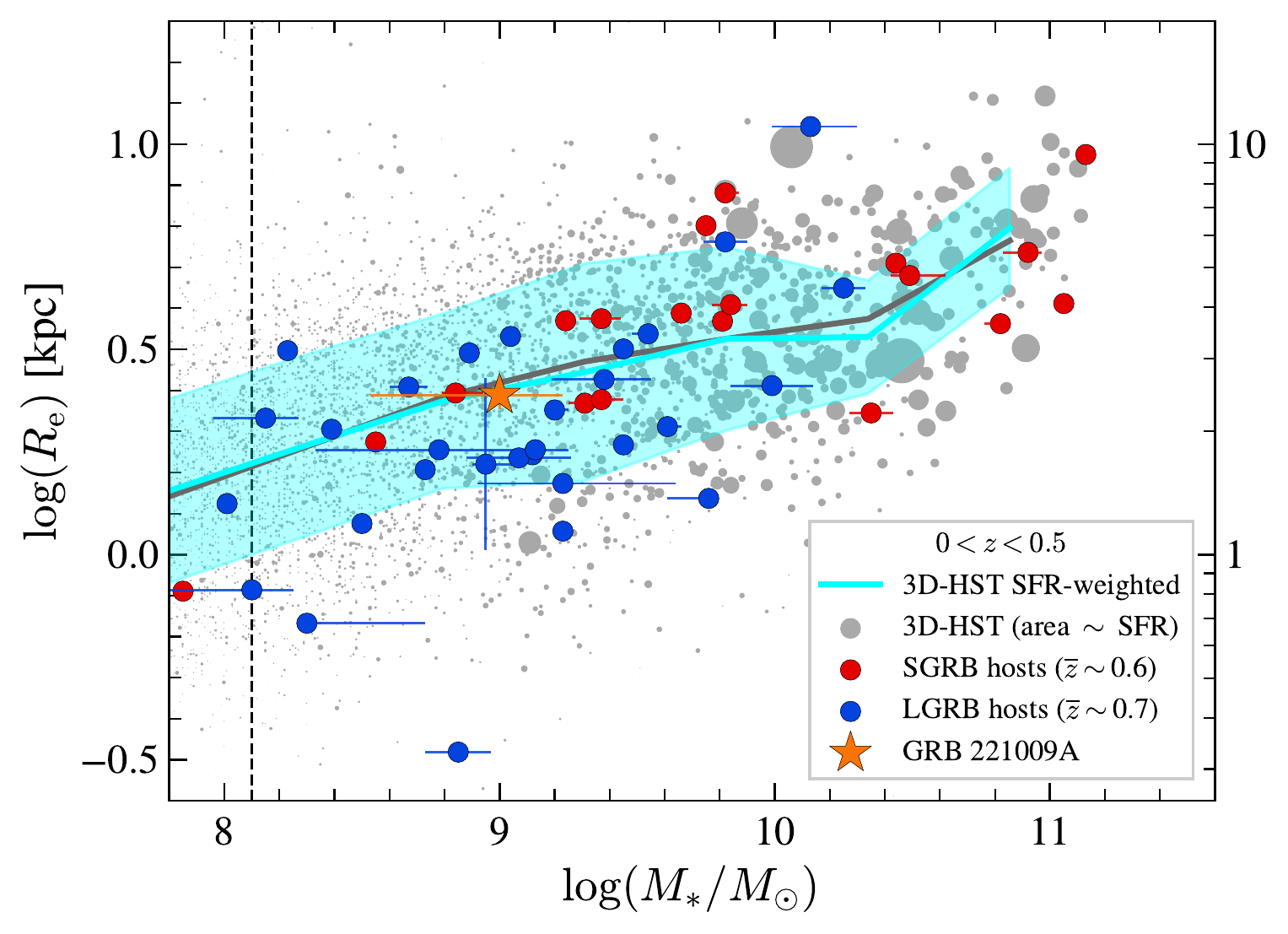}
     \caption{Effective radius as a function of F160W magnitude (left) and against stellar mass (right) for GRB hosts and 3D-HST star-forming galaxies at $0<z<0.5$. The red circles are a population of short GRB hosts at $z<1$ from \cite{2022ApJ...940...56F} and \cite{2022ApJ...940...57N}, and blue circles a population of long GRB hosts at $z<1$ from \cite{2014ApJ...789...23K} and \cite{2016ApJ...817..144B}. The GRB~221009A host galaxy is visible as an orange star marker. The SFR-weighted median of the 3D-HST population and its 1$\sigma$ uncertainty are shown as a cyan line and a surrounding cyan region.}
     \label{fig:mass-size}
\end{figure}

Our \galfitm model also provides us with an estimate of the galaxy size. We determine an effective radius of $R_e = 2.45 \pm 0.20$~kpc for the F160W filter. In addition, using a mass-to-light ratio derived from the star-forming galaxies of the COSMOS2015 catalog \citep{Laigle2016} at $z_{\mathrm{GRB}} \pm 0.1$ and the \galfitm F160W magnitude corrected for extinction, we estimate a stellar mass of $\lmass = 9.00^{-0.47}_{+0.23}$. The comparison with a star-forming population of the 3D-HST survey \citep{Skelton2014,vanderWel2014,Momcheva2016} at a similar redshift suggests that the host size is typical for this epoch (left and right panels of Figure~\ref{fig:mass-size}). Although, we note that this host galaxy seems to populate the lower part of the SFR-weighted median of field galaxies, as previously observed for long GRB host galaxies up to $z \sim 2$ \citep{2014ApJ...789...23K, Schneider2022}. Compared to the populations of short and long GRB hosts (right panel of Figure~\ref{fig:mass-size}), the size of the GRB221009A host appears to be more similar to long GRB hosts than to short GRB hosts that populate larger and more massive galaxies.
For F160W, our model returned a \sersic index of $n = 1.71 \pm 0.18$ and an axis ratio of $b/a = 0.22 \pm 0.01$, in agreement with the apparent edge-on, disc-like morphology observed in the WFC3/IR images. This morphology seems slightly more unusual amongst GRB hosts \citep{lyman17}, although some local events do appear in such galaxies.
These diagnostics suggest that the host of GRB 221009A is not especially unusual amongst the hosts of either long or short-duration GRBs.

The location within the host galaxy and the host galaxy characteristics appear to be very typical of long GRBs. 
This argues that the Galactic environment of GRB\,221009A is not the cause of its extreme properties, and, in particular, there is no evidence that it was
spawned in a very low metallicity system.

\section{Discussion}
\subsection{Implications for afterglow models} \label{sec:afterglow}
\label{afterglow_disc}
GRB afterglows are well described by the dynamics of a relativistic shell colliding with an external medium and synchrotron emission from the associated shocks \citep[e.g.][]{sari1998}. The physics of the broadband spectral and temporal behaviour of the afterglow are linked through the various closure relations that depend on the spectral regime, observed time, and density of the external medium \citep{granot2002}. Changes in the observed spectral and temporal power law indices are useful diagnostic tools, when compared with the expectation from GRB afterglow relations, they can be used to identify the spectral regime of the observed afterglow components and determine the external medium density profile -- typically a uniform interstellar medium (ISM), or stellar wind.

For GRB 221009A, the spectral index changes from $\beta\sim0.8\pm0.1$ at X-ray frequencies to $\beta \sim0.3\pm0.1$ at NIR. This change is consistent with that expected from the cooling break ($\nu_c$), $\beta_{\nu>\nu_c} - \beta_{\nu<\nu_c} = \Delta\beta=0.5$, where above the cooling break frequency, synchrotron electrons will lose energy significantly.
For $\nu>\nu_c$, then $\beta=p/2$, where $p$ is the power law index of the Lorentz factor distribution for the shock accelerated electrons that produce the synchrotron emission. Given a spectral index of $\beta = 0.8$ we find $p=1.6$ for $\nu>\nu_c$; where for GRB afterglows, the value of $p$ typically lies in the range $2\leq p\leq3$, however, a value of $p<2$ is not unusual \citep[e.g.,][]{higgins2019}.
Similarly, for $\nu<\nu_c$, then $\beta=(p-1)/2$, and for $\beta=0.3$ we find a consistent, $p=1.6$.
However, the steep temporal decline of the afterglow (where  $F_\nu(t)\propto\nu^{-\beta} t^{-\alpha}$), with $\alpha\sim1.67$ at X-ray frequencies \citep{williams2023}, and $\alpha\sim 1.44$ at optical/NIR \citep{shrestha2023} is not consistent with the expected decline rate, given these spectral regimes and $p<2$ using standard closure relations.
To resolve this, we invoke a very early jet break without significant lateral spreading, {that steepens the temporal index by $3/4$, the temporal index for an ISM medium with $p<2$ is then \citep{gao2013},
\begin{eqnarray}
    \alpha_{1} =& \frac{3(p+6)}{16}, \quad  &(\nu<\nu_c)\\
    \alpha_{2} =& \frac{3p+22}{16}, \quad &(\nu>\nu_c).
\end{eqnarray}
These relations provide decline indices of $\alpha_2\sim1.675$ and $\alpha_1\sim1.425$ and are similar to the observed afterglow decline rates at both X-ray and optical/NIR frequencies.
This result is consistent with the ``no observed supernova" scenario described in \cite{shrestha2023}, where the lightcurve is dominated by emission from the GRB afterglow only with no significant supernova contribution, and consistent with the lack of any supernova features in the spectra at $>10$\, days.

For the cooling break to be found between optical and X-ray frequencies for the duration of the observed afterglow, we require an energy on the order of $10^{54}$ erg, in a uniform ISM with $n\sim 1$\,cm$^{-3}$ \citep[see e.g.,][]{kann23}, and fixing microphysical parameters (that describe the fraction of the shock energy that goes into the magnetic field, $\varepsilon_B$, and the accelerated electrons, $\varepsilon_e$) to $\varepsilon_B = 0.01$ and $\varepsilon_e=0.1$, gives $\nu_c \sim 6\times10^{15}$\,Hz, $\nu_a \sim 40$ GHz, and $\nu_m \sim 3$\,GHz at $\sim0.5$\,days and $\nu_c \sim 10^{15}$\,Hz, $\nu_a \sim 4$\,GHz and $\nu_m \sim 2$\,MHz at 13 days, where $\nu_m$ is the characteristic synchrotron frequency, and $\nu_a$ is the self-absorption frequency.
The spectral and temporal peak emission for the synchrotron process is at the characteristic synchrotron frequency unless $\nu_a>\nu_m$, then the spectra will peak at the self-absorption frequency.
In the $p<2$ regime, where $\nu_m<\nu_a<\nu_c$\,$(\nu_a<\nu_m<\nu_c)$, then $\nu_a \propto t^{-(3p+26)/[8(p+4)]}\,(t^{-[9(p-2)]/[16(p-1)]})$, and for $p=1.6$ we have $\nu_a\propto t^{-11/16}\,(t^{3/8})$, and decreasing(increasing) with time.
Emission below the synchrotron self-absorption frequency will be suppressed and post-jet-break will evolve as $t^{1/2}$ where $\nu_m<\nu<\nu_a$ \citep[and $t^{-(14-5p)/[16(p-1)]}$ or $t^{-5/8}$ with $p=1.6$, where  $\nu<\nu_a<\nu_m$, see e.g.,][]{gao2013}.

These spectral break frequencies are approximately consistent with the extrapolated join in the optical to X-ray spectrum (see Fig. \ref{fig:spec_fits}), and, although the self-absorption frequency is a factor of a few above that seen in early radio observations \citep{laskar2023}, the radio peak frequency and spectral behaviour is $\sim$consistent if we assume that our $\nu_a$ is underestimated.
An estimate of the maximum flux for this model ($F_{\rm max} \sim 10$\,Jy at 0.5\,days, and noting that as $\nu_m<\nu_a$ the spectral peak at $\nu_a\sim40$\,GHz results in a suppressed peak flux, $F\sim400$\,mJy) and, following the closure relations in \cite{gao2013}, the flux density from a post-jet break and $p<2$ model, evolves with time as $t^{-1.425}$ for $p=1.6$ and $\nu_m<\nu_a<\nu$.
The predicted light curve has a slightly more rapid decline than that seen in radio observations \citep{laskar2023}, however, the reverse shock can peak at radio frequencies and may contribute to this low-frequency emission.
Alternatively, post jet break the emission from higher latitudes given a so-called ``structured jet", where significant energy extends beyond the highly collimated jet core region, can contribute to the lightcurve \citep{lamb2021} e.g., see \cite{sato2022,oconnor23} who use structured jet models to explain the irregular temporal behaviour of the afterglow for this GRB.
However, it is beyond the scope of this work to precisely model and fit the full afterglow light curve.

Using the estimate above for the energy and ambient density, and requiring a jet-break at $<0.03$\,days, the time at which the lightcurve temporal index will steepen by $3/4$, we find a very narrow jet with a half-opening angle, $\theta_j<0.02$\,rads or $\sim1.15$\,deg, and a Lorentz factor, $\Gamma\sim50$ at the jet break time.
The transition to the Newtonian regime for a decelerating blastwave occurs when $\Gamma(t) \sim \sqrt{2}$.
Where $\Gamma\sim50$ at 0.03 days, then {as $\Gamma(t)\propto t^{-3/8}$, the transition to the Newtonian regime will occur on the order of $\sim1$ year.

Without the NIR spectral index of $\beta\sim0.3\pm0.1$, it is tempting to assume a wind-like medium for the afterglow model \citep[e.g.,][]{ren2022, laskar2023}.
As noted above, the change in the spectral index from higher energies to the NIR, $\Delta\beta\sim0.5$, indicates the cooling break, $\nu_c$, between X-ray and optical frequencies.
For a wind-like medium, the temporal decline at $\nu>\nu_c$ is shallower than the decline at $\nu<\nu_c$.
However, the observed lightcurve has a shallower decline at optical/NIR than at X-ray frequencies, which rules out the cooling break as the origin of the observed change in the spectral index if we invoke a wind-like medium \citep[however, see the discussion in][for arguments in favour of a wind-like environment]{laskar2023}.
Therefore, such a model requires that X-ray to NIR occupy the same spectral regime, $\nu_m<\nu<\nu_c$, giving $p\sim2.6$ and a common temporal index at X-ray energies through optical to NIR, and contrary to the observed lightcurve evolution which differs between the optical and X-ray regimes.
The spectral index at $\nu<\nu_m$ is $\beta = -1/3$, with a spectral peak that evolves with time as $\nu_m\propto t^{-3/2}$.
If the recent passage of this spectral break was the cause of the $\beta\sim0.2$ spectral index in the X-shooter spectrum at $\sim0.5$\,days, then a chromatic break in the optical to NIR lightcurve, from $t^{0}$ to $t^{-1.7}$ for $p=2.6$, would need to be present before this time -- such a change is not seen.
Additionally, the $\nu_m$ spectral break is expected to be relatively sharp, unlike the cooling break \citep[see][]{uhm2014}, and should be well below the wavelength range of the \jwst\ spectra at $\sim13$ days, making the observed $\beta\sim 0.4$ at NIR difficult to explain via a wind-medium model.
We, therefore, favour an early jet break, $p<2$, uniform ISM environment, and $\nu_m<\nu_a<\nu_c$ spectral order to explain the afterglow.

\subsection{Implications for the progenitor of GRB 221009A}
A striking result is an apparent absence of
any supernova emission in GRB 221009A. We note that this result conflicts with claims on supernovae to date \citep{fulton23}, which are based on substantially more complete photometric coverage but lack the high S/N ratio spectral information presented here. In particular, in the analysis of \cite{fulton23}, the supernova should contribute essentially no light to the afterglow+SN combination at $\sim 0.5$ days, but $\sim 20-30$\% at the time of the \jwst\ observations, and $>$50\% at the time of the \hst\ observations. Such a supernova should be visible as a marked change in the F625W-F098M colour (or as an excess visible in the NIRSPEC spectrum). This is not the case. In part, this may reflect assumptions about the afterglow's underlying spectral and temporal behaviour. Isolating any supernova component within this burst is not straightforward, given the issues associated with high foreground extinction and crowding.

The lack of any associated supernova would be 
surprising within this GRB. Although there have been several long GRBs seen without apparently associated supernovae \citep{fynbo06,galyam06,dellavalle06}, in most cases, these have now been suggested to arise from compact object mergers \citep{gehrels06,rastinejad22}, with evidence for kilonovae in several cases \citep{rastinejad22,troja22,yang22,jin15,yang15}. 
GRB 221009A may belong to this class. However, the energetics of 
GRB 221009A lie substantially beyond any seen in other merger-origin  GRBs. For example, the recent
GRB 211211A has $E_{\rm iso} = 7 \times 10^{51}$\,erg \citep{gompertz23,mei22}, almost three orders of magnitude less energetic that of GRB\,221009A. The most energetic short GRBs at higher redshift also have $E_{\rm iso} \sim 5 \times 10^{52}$
erg \citep{fong15}. 

An alternative explanation for supernova-less long GRBs is that they arise from direct collapse to black holes in which insufficient material is launched into an associated shock to power a successful supernova \citep{fynbo06}. Such events may arise in GRBs from very massive stars. If such stars have a substantial energy reservoir for the GRB, one may get a very luminous GRB without an associated supernova. 
Very massive stars may be more common at lower metallicity. At least, stars which retain sufficient
mass at later times are expected to occur more frequently at metallicities where wind-driven mass loss becomes less important \citep{heger03,fryer19}. 

Finally, and perhaps most likely, GRB supernovae have a modest range of luminosities, as well as 
evolution timescales and colours \citep{Cano17}. Given the difficulties in isolating the GRB afterglow, host galaxy and supernova light, it is plausible that an event somewhat less luminous than SN~1998bw (and perhaps somewhat faster evolving or bluer) could have evaded detection in our observations. Ultimately, once deep images are available for image subtraction, the \hst\ photometry should allow S/N $>$100 measurements of the spectral shape. We may then expect to uncover the associated supernova light. 

\section{Conclusions}
We have presented a series of high S/N ratio measurements of the spectral shape of the optical to mid-IR afterglow
of GRB 221009A. These data, at high confidence, demonstrate that the optical/IR shape is not the same as that seen in the X-ray or as the X-ray to optical index. This suggests that the two regimes lie on different branches of the synchrotron spectrum. The separation in the spectral slopes of $\Delta \beta \sim 0.5$ make the difference most likely due to the cooling break. However, tensions with other multi-wavelength data remain, and these do not have straightforward solutions \citep[e.g.][]{laskar2023,williams2023}. 

The optical to mid-IR (0.6-12 micron) spectral energy distribution shows little evidence for variability from early to late times (0.5 to 55 days). The lack of variability implies, at most, modest contributions from supernova emission at these epochs or that the supernova emission peaks outside of the wavelength range covered (e.g. to the blue). The challenges of high foreground extinction and a bright afterglow will continue to make studying the supernova associated with GRB 221009A challenging. Still, accurate host subtraction, combined with the collation of the extensive, coherent data obtained for GRB 221009A, should enable much better constraints. It is unlikely (although possible in the absence of clear-cut evidence for a supernova) that GRB 221009A is created through a compact object merger. 

The burst's environment (i.e. its host galaxy) appears very broadly typical of the long GRB population. There is no evidence of an unusual galaxy or location within the host. This, in turn, implies that the environment in which GRB 221009A formed is comparable to those of other long GRBs at low redshift. For example, it does not match where we may expect to locate very low metallicity gas, or especially massive stars. Hence, the extreme properties of the burst are likely not linked to an extreme and unusual environment.

\begin{acknowledgments}
We thank the referee for a prompt and constructive report on a paper which is clearly longer than the average ApJ letter.
We thank the staff of STScI for their work in rapidly scheduling approving and these observations, in particular Katey Alatalo, Claus Leitherer, Alison Vick, William Januszweski, Greg Sloan and Patrick Ogle. 

This research is based on observations made with the NASA/ESA {\em Hubble Space Telescope} obtained from the Space Telescope Science Institute, which is operated by the Association of Universities for Research in Astronomy, Inc., under NASA contract NAS 5–26555. These observations are associated with program(s) 17264. This work is based in part on observations made with the NASA/ESA/CSA {\em James Webb Space Telescope}. The data were obtained from the Mikulski Archive for Space Telescopes at the Space Telescope Science Institute, which is operated by the Association of Universities for Research in Astronomy, Inc., under NASA contract NAS 5-03127 for \jwst. These observations are associated with program \#2782. Partly based on observations made with the Gran Telescopio Canarias (GTC), installed at the Spanish Observatorio del Roque de los Muchachos of the Instituto de Astrofísica de Canarias, on the island of La Palma. Partly based on observations carried out under project numbers S22BC with the IRAM NOEMA Interferometer. IRAM is supported by INSU/CNRS (France), MPG (Germany) and IGN (Spain). Partly based on observations collected at the European Southern Observatory under ESO programme 110.24CF (PI Tanvir). Based on observations made with the Italian Telescopio Nazionale Galileo (TNG) operated on the island of La Palma by the Fundación Galileo Galilei of the INAF (Istituto Nazionale di Astrofisica) at the Spanish Observatorio del Roque de los Muchachos of the Instituto de Astrofisica de Canarias.

AJL, DBM and NRT are supported by the European Research Council (ERC) under the European Union’s Horizon 2020 research and innovation programme (grant agreement No.~725246).
GPL is supported by a Royal Society Dorothy Hodgkin Fellowship (grant numbers DHF-R1-221175 and DHF-ERE-221005).
JH and LI were supported by a VILLUM FONDEN Investigator grant to JH (project number 16599).
BDM acknowledges support from the National Science Foundation (grant \#AST-2002577).
JPUF acknowledges support from the Carlsberg Foundation. The Cosmic Dawn Center 
(DAWN) is funded by the Danish National Research Foundation under grant No.140.
DAK acknowledges the support by the State of Hessen within the Research Cluster ELEMENTS (Project ID 500/10.006).
RB, MGB, SC, PDA, MF, AM and SP acknowledge funding from the Italian Space Agency, contract ASI/INAF n. I/004/11/4.
PDA acknowledges support from PRIN-MIUR 2017 (grant 20179ZF5KS).
JFAF acknowledges support from the Spanish Ministerio de Ciencia, Innovaci\'on y Universidades through the grant PRE2018-086507.
\end{acknowledgments}

\vspace{5mm}
\facilities{JWST(NIRSPEC/MIRI), HST (WFC3), VLT, GTC, TNG, NOEMA}

\software{astropy \citep{2013A&A...558A..33A,2018AJ....156..123A},  
          emcee \citep{2013PASP..125..306F}, GALFIT \citep{Peng2010}}

Some/all of the data presented in this paper were obtained from the Mikulski Archive for Space Telescopes (MAST) at the Space Telescope Science Institute. The specific observations analyzed can be accessed via \dataset[DOI: 10.17909/cs82-x148]{https://doi.org/10.17909/cs82-x148} and \dataset[DOI: 10.17909/zr2q-sx52]{https://doi.org/10.17909/zr2q-sx52}.

\bibliographystyle{aasjournal}
\bibliography{refs}

\end{document}